\documentclass[a4paper,11pt]{article}
\usepackage{pos}

\title{Nobeyama Cygnus-X Survey: Physical Properties of C$^{18}$O clumps in DR-6(W), DR-9 and DR-13S regions*\footnote{*As part of the thesis to be submitted by Toledano–Ju\'arez as a partial fulfillment for the requirements of Ph. D. Degree in Physics, Doctorado en Ciencias (F\'isica), CUCEI, Universidad de Guadalajara}}

 \ShortTitle{Clumps in DR-6W, DR-9, and DR13S regions}

\author[a]{I. Toledano--Ju\'arez}
\author*[b]{E. de la Fuente}
\author[c]{K. Kawata}
\author[e]{M. A. Trinidad}
\author[e]{D. Tafoya}
\author[f]{M. Yamagishi}
\author[g]{S. Takekawa}
\author[b]{M. Ohnishi}
\author[h]{A. Nishimura}
\author[h]{S. Kato}
\author[h]{T. Sako}
\author[h]{M. Takita}
\author[i]{R. K. Yadav}

\affiliation[a]{Doctorado en Ciencias en F\'{i}sica, CUCEI, Universidad de Guadalajara,\\
  Blvd. Marcelino Garc\'{i}a Barrag\'an 1420, 44430, Guadalajara, Jalisco, M\'exico}  

\affiliation[b]{Departamento de F\'{i}sica, CUCEI, Universidad de Guadalajara,\\
  Blvd. Marcelino Garc\'{i}a Barrag\'an 1420, 44430, Guadalajara, Jalisco, M\'exico}

\affiliation[c]{Institute for Cosmic Ray Research, University of Tokyo\\
  Kashiwa 277-8582, Japan} 

\affiliation[d]{Departamento de Astronom\'{i}a, Universidad de Guanajuato\\
  Apartado Postal 144, 36000, Guanajuato, Guanajuato, M\'exico}  

\affiliation[e]{Department of Space, Earth, and Enviroment, Chalmers University of Technology,\\
  Onsala Space Observatory, 439 92 , Sweden}

\affiliation[f]{Institute of Astronomy, Graduate School of Science, The University of Tokyo\\
  2-21-1 Osawa, Mitaka, Tokyo 181-0015, Japan}    

\affiliation[g]{Department of Applied Physics, Faculty of Engineering, Kanagawa University\\
  3-27-1 Rokkakubashi, Kanagawa-ku, Yokohama, Kanagawa 221-8686, Japan}  
\affiliation[h]{Nobeyama Radio Observatory, National Astronomical Observatory of Japan, National Institutes of Natural Sciences 462-2 Nobeyama, Minamimaki, Minamisaku, Nagano 384-1305, Japan}  
\affiliation[i]{National Astronomical Research Institute of Thailand (Public Organization)\\
  260 Moo 4, T. Donkaew, A. Maerim, Chiangmai, 50180, Thailand} 

\emailAdd{ivan.toledano9284@alumnos.udg.mx}

\abstract{\textbf{Abstract}. Cygnus-X is considered a region of interest for high-energy astrophysics, since the Cygnus OB2 association has been confirmed as a PeVatron in the Cygnus cocoon. In this research note, we present new high-resolution (16'') $^{12,13}$CO(J=1$\rightarrow$0) and C$^{18}$O (J=1$\rightarrow$0) observations obtained with the Nobeyama 45-m radiotelescope, to complement the Nobeyama Cygnus-X Survey. We discovered 19 new C$^{18}$O clumps associated with the star-forming regions DR-6W, DR-9, and DR13S. We present the physical parameters of these clumps, which are consistent with the neighboring covered regions. We confirm the clumpy nature of these regions and of a filament located between DR6 and DR6W. These results strongly suggest that star formation occurs in these regions with clumps of sizes $\sim$10$^{-1}$ pc, masses $\sim$10$^2$ M$_\odot$, and H$_2$ densities of $\sim$10$^4$ cm$^{-3}$.

}

\ConferenceLogo{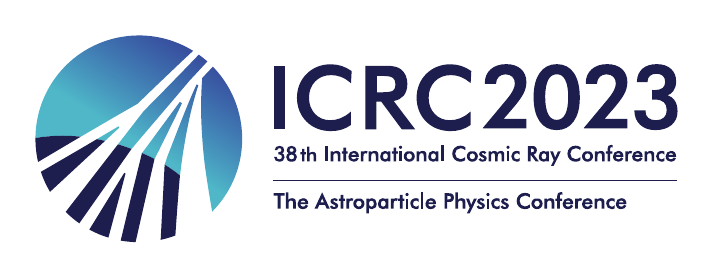}

\FullConference{%
38th International Cosmic Ray Conference (ICRC2023)\\
  26 July - 3 August, 2023\\
  Nagoya, Japan}

\begin{document}
\maketitle

\section{Introduction}

PeVatrons have become a hot topic in high energy astrophysics thanks to observations by the HAWC (e.g \cite{Abeysekara2023}), Tibet-AS$\gamma$ (e.g. \cite{Amenomori2019}) and LHAASO (e.g. \cite{Ma2022}) in 2021 (\cite{Abeysekara2021,Amenomori2021a,Cao2021}). Currently under debate, PeVatrons can be defined as natural occurring cosmic-ray particle accelerators capable of accelerating particles to PeV energies, emitting gamma rays with energies above 100 TeV (e.g. \cite{Gabici2019,delaFuente2023a} and references therein). The Cygnus region holds significant importance in gamma-ray astrophysics, primarily due to the presence of three of the first twelve reported PeVatron candidates by \cite{Cao2021}: LHAASO J2108+5157, the Dragonfly Nebula associated with PSR J2021+361, and the massive star cluster Cygnus-OB2. In fact, HAWC (see \cite{Abeysekara2021} and references therein) confirmed Cygnus-OB2 as the PeVatron associated with the Cygnus cocoon region, originally discovered in the Cygnus-X region by \textit{Fermi}-LAT (\cite{Ackermann2011}). This was the first time that a massive star cluster was confirmed as an accelerator for cosmic-ray particles (see the theory of \cite{Aharonian2019}). In this context, the following question naturally arises: Does Cygnus-OB2 stand alone as the sole PeVatron in the vicinity of the Cygnus cocoon, responsible for the gamma-ray emission from both the cocoon and its surroundings?

\begin{figure*}[!ht]
\begin{center}
\includegraphics[width=0.49\columnwidth]{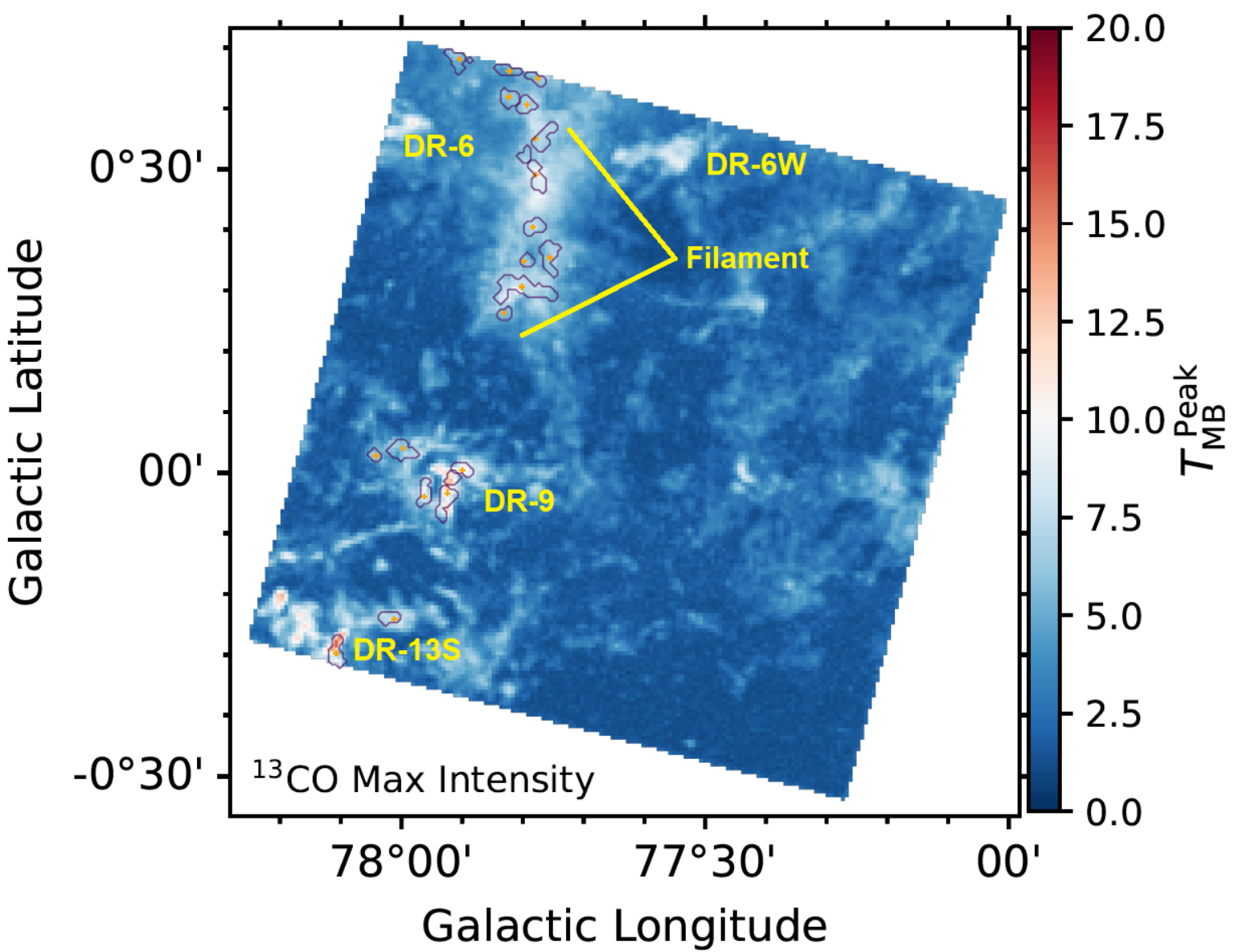}
\includegraphics[width=0.49\columnwidth]{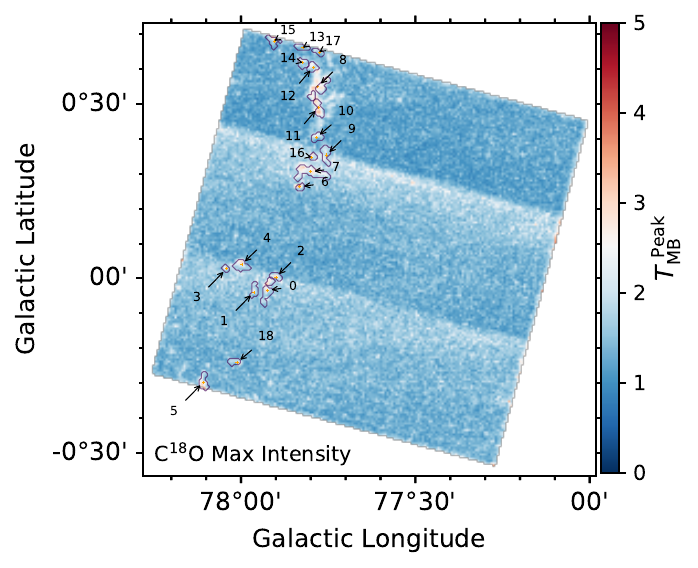}
\includegraphics[width=0.45\columnwidth]{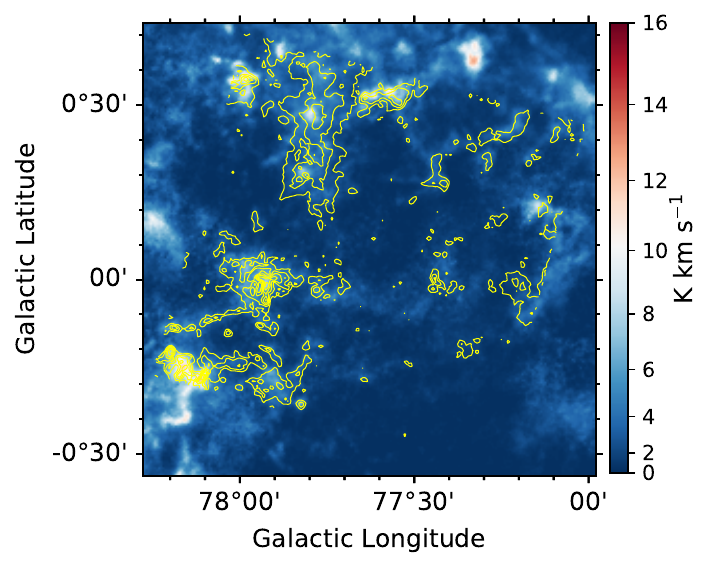}
\includegraphics[width=0.45\columnwidth]{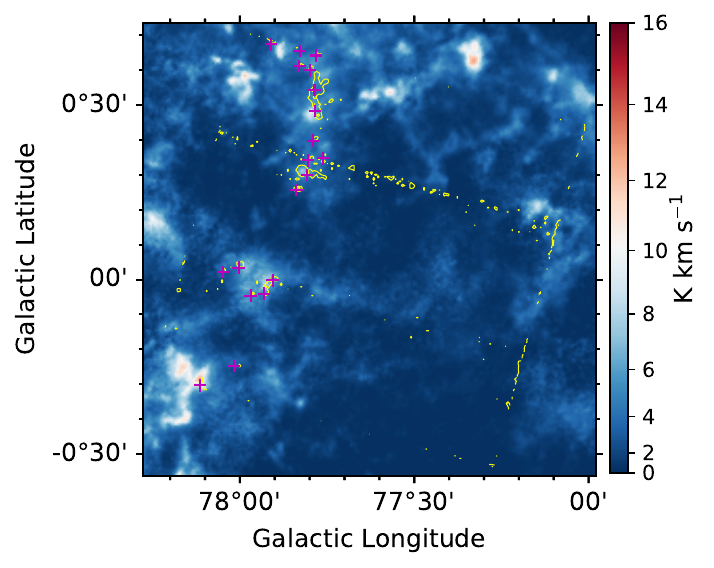}
\end{center}
\caption{\textbf{Top}: Nobeyama $^{13}$CO (left panel) and C$^{18}$O (right panel) observations of the Cygnus-X molecular cloud that covers the DR6W, DR9, and DR13S regions. The clumps are shown as contours on the $^{13}$CO map and labeled in C$^{18}$O. Clumps 0 to 4 are related to DR-9, clump 5 is related to DR13-S, clump 18 is isolated, and the rest of clumps are associated with the filament between DR-6 and DR-6W  \textbf{Bottom}: The lower part of the figure displays FCRAO observations of $^{13}$CO (left) and C$^{18}$O (right) represented in colors, while the corresponding Nobeyama emission is shown as contours. The contour levels for $^{13}$CO are set at [-4, 4, 6, 8, 10, 12, 14] times the root mean square (rms) of 1.0 K, and for C$^{18}$O, they are set at [-5, 5, 7, 9, 11] times the rms of 0.45 K. The magenta crosses show the positions of the reported clumps labeled in the top right panel.}   
\label{fig:c18o_mom8_clumps}
\end{figure*}

\begin{figure*}[!ht]
\begin{center}
\includegraphics[width=0.49\textwidth]{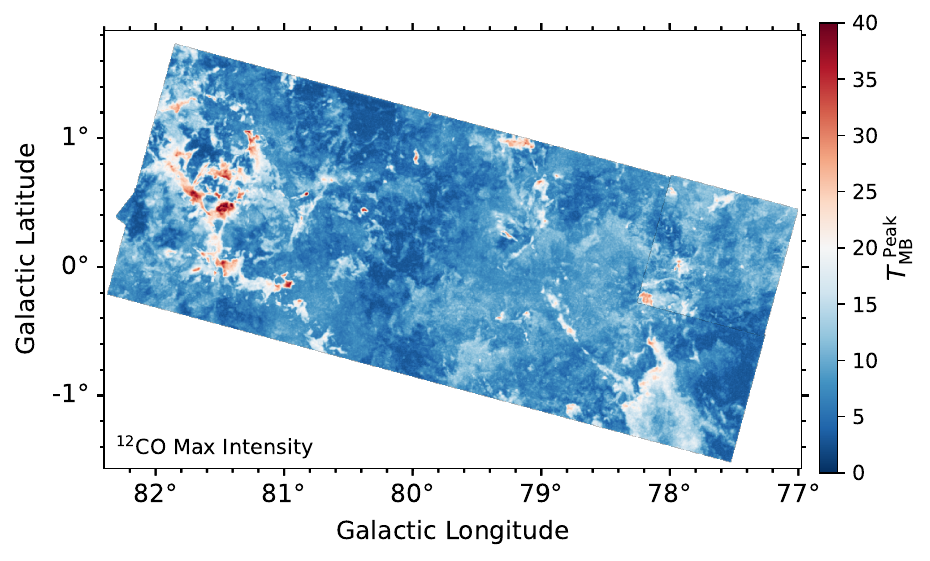}
\includegraphics[width=0.49\textwidth]{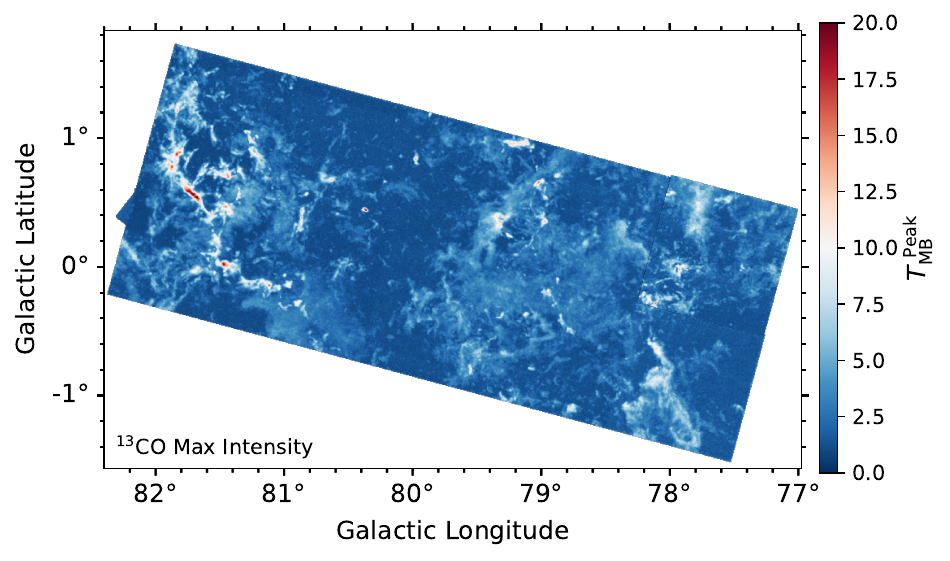}
\includegraphics[width=0.49\textwidth]{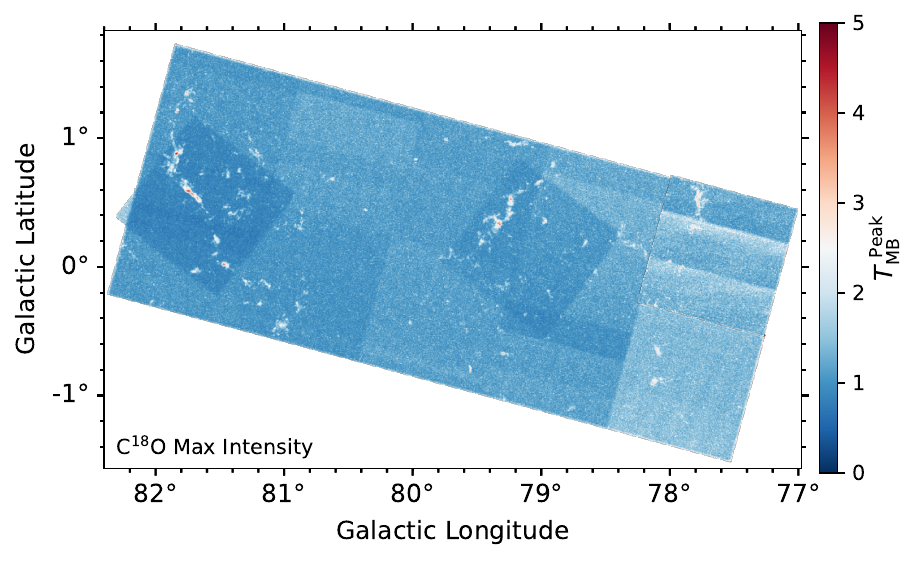}
\includegraphics[width=0.49\textwidth]{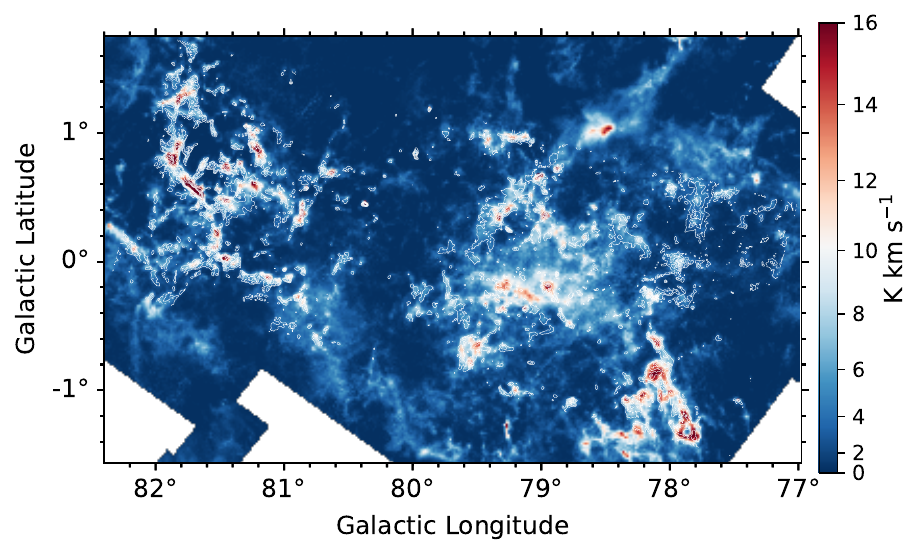}

\end{center}
\caption{Max intensity maps of Nobeyama observations at $^{12}$CO(J=1$\rightarrow$0) on the top-left, $^{13}$CO(J=1$\rightarrow$0) on the top-right, and C$^{18}$O(J=1$\rightarrow$0) on the bottom-left, for the whole Nobeyama Cygnus-X Survey. Comparing our Fig.~\ref{fig:c18o_mom8_clumps} with Figs. 1 and 2 of \cite{Takekoshi2019}, the new square-degree region is on the top right. FCRAO $^{13}$CO(J=1$\rightarrow$0) in color scale with Nobeyama $^{13}$CO(J=1$\rightarrow$0) overlaid as contours (same as Fig.~\ref{fig:c18o_mom8_clumps}) are presented in the bottom right panel. FCRAO emission covers a wider field in Cygnus-X than in the Nobeyama Cygnus-X Survey.}
\label{fig:mosaic_12co_13co_c18o}
\end{figure*}

A major unresolved inquiry in the study of PeVatrons, still subject to debate, revolves around determining the nature of their emission: whether it is hadronic, involving protons, or leptonic, involving electrons. Investigating the presence of molecular gas in PeVatrons is of significant importance, as it contributes to comprehending the hadronic origin of gamma-ray emission resulting from neutral-pion decays, which stem from the interaction of cosmic rays with molecular clouds (e.g. the enigmatic Pevatron candidate LHAASO J2108+5157 in Cygnus-OB7; \cite{delaFuente2023b} and references therein). For a comprehensive overview of the Cygnus region, see \cite{Reipurth2008}. other detailed studies of the Cygnus-X region have been conducted by \cite{Schneider2006,Schneider2011} and references therein. Distances between 1.5 and 2 kpc for certain Cygnus OB stellar associations (approximately 1.7 kpc for Cygnus OB2) are considered by these authors. Regarding observational data, low-resolution Kölner Observatorium für Submm-Astronomie (KOSMA) observations of $^{13}$CO(J=2$\rightarrow$1), $^{13}$CO(J=3$\rightarrow$2), $^{12}$CO(J=3$\rightarrow$2), as well as Five College Radio Astronomical Observatory (FCRAO) observations of $^{13}$CO(J=1$\rightarrow$0) and C$^{18}$O(J=1$\rightarrow$0) covering several degrees in the Cygnus-X region have been presented by \cite{Schneider2006,Schneider2011}.

The Nobeyama Cygnus--X Survey, with observations of $^{12,13}$CO(J=1$\rightarrow$0) and C$^{18}$O(J=1$\rightarrow$0) at an angular resolution of 16$^{\prime\prime}$, was started by \cite{Takekoshi2019,Yamagishi2018} studying the star--forming regions DR 21, DR 15, DR13S and their surroundings. In their study, these authors presented a mosaic comprised of nine individual images, each image covering an area of one square degree (see their Figs. 1, 2 and 3, respectively). However, it is important to note that the western part of the stars-forming regions DR13-S, DR6, DR6-W, and DR9 were excluded from this mosaic. Instead, these regions were observed separately at low-resolution using the KOSMA and FCRAO instruments. Remarkably, within the Nobeyama missing field, the Tibet-AS$\gamma$ observatory has reported the detection of gamma-ray emission at sub-PeV energies (\cite{Amenomori2021b}). This result emphasizes the need for a comprehensive investigation of the molecular gas in this particular region. The study C$^{18}$O clumps is essential for a better understanding of the underlying processes. The molecular gas serves as a tracer for the cosmic rays responsible for the observed gamma-ray emission resulting from the decay of neutral pions (e.g., LHAASO J2107 + 5158 \cite{delaFuente2023b}). Moreover, the presence of young embedded star clusters in regions of massive star formation (potentially associated with PeVatrons) can be identified by the presence of C$^{18}$O clumps (e.g., \cite{Saito2007}). From this study, a pertinent question arises: Is Cygnus-OB2 the unique PeVatron responsible for all the observed gamma-ray emissions (leptonic and/or hadronic) within the entire Cygnus-X star-forming region? Given the significance of Cygnus-X, it warrants comprehensive investigations using high-resolution observations, such as those facilitated by the Nobeyama Radio Observatory (NRO) telescopes. 

To complement the findings of \cite{Takekoshi2019} and \cite{Yamagishi2018}, and to facilitate a comparison with \cite{Schneider2006} and \cite{Schneider2011}, we present our high-resolution (16'') observations obtained from the NRO 45m radio telescope. These observations cover the missing and complementary one-deg$^2$ region discussed earlier. Consistent with the methodology of \cite{Takekoshi2019}, we have determined the physical parameters of the clumps, aligned with similar regions examined in their previous studies. Section \ref{sec:observations} provides a detailed account of the observations and procedures used for clump identification. Subsequently, in Section \ref{sec:res_disc}, we present the results and engage in a comprehensive discussion. Finally, our conclusions are summarized in Section \ref{sec:conclusions}.

\section{Observations and C$^{18}$O clump identification}
\label{sec:observations}

The $^{12,13}$CO(J=1$\rightarrow$0) and C$^{18}$O(J=1$\rightarrow$0) observations were performed in the first quarter of 2023 using the FOREST receiver and covering a local standard of rest (LSR) velocity range of -80 to 80 km s$^{-1}$. The field of view is one square degree with an angular resolution of 16$''$ and an integration time of 870 minutes for 660 scans. Details of setup and data reduction are described in \cite{delaFuente2023b}. %The $^{12,13}$CO(J=1$\rightarrow$0) and C$^{18}$O(J=1$\rightarrow$0) observations were conducted during the first quarter of 2023 using the FOREST receiver. The observations covered a local standard of rest (LSR) velocity range spanning from $-$80 to 80 km s$^{-1}$. The field of view covered one square degree and was observed with an angular resolution of 16$''$. The integration time for the observations involved 870 minutes spread across 660 scans. For further details on the setup and data reduction procedures, refer to \cite{delaFuente2023}. 
The data reduction process followed standard procedures using NOSTAR software\footnote{\url{https://www.nro.nao.ac.jp/~nro45mrt/html/obs/otf/export-e.html}}. To be able to compare the resulting data cubes with those presented in \cite{Takekoshi2019}, we convolved our cube using a 22.7$^{''}$ pixel grid and maintained a velocity resolution of 0.25 km s$^{-1}$. For calibration purposes, we considered the main beam efficiencies $\eta_{\rm MB}$ of 38.9\% and 39.9\% at 115 and 110 GHz, respectively, allowing us to obtain a main beam brightness temperature scale. The average root mean square ($T_{\rm rms}$) noise per channel was approximately 0.40 K, 0.17 K, and 0.46 K, for the $^{12}$CO, $^{13}$CO and C$^{18}$O data cubes, respectively.

%To compare our data cube with those reported in \cite{Takekoshi2019,Yamagishi2018}, we IVAN.

%To detect C$^{18}$O clumps, we first dentified the Downes \& Rinehart (DR) star-forming  regions (\cite{Downes1966}) in the field and then, following \cite{Takekoshi2019}, \textbf{we used Python astrodendro\footnote{\url{http://www.dendrograms.org/}} package}. 

In the initial stage, we identified the Downes \& Rinehart (DR) star-forming regions as described in \cite{Downes1966} within the observed field. Subsequently, following the approach employed by \cite{Takekoshi2019}, we detected clumps using the Python package astrodendro\footnote{\url{http://www.dendrograms.org/}}. The identification process involved setting specific parameters, $T_{\rm min} = 3 T_{\rm rms}$, $T_{\rm delta} = 2 T_{\rm rms}$, and $n_{\rm vox} = 15$, the latter being the minimum number of voxel pixels used to identify a cluster in the data cube. Identified clumps that overlap or totally engulf a smaller clump with a similar LSR velocity are treated as false detection and are discarded. For each C$^{18}$O clump, we extracted the spectra and applied a 2-D Gaussian fitting method to obtain the peak main-beam temperature $T_{\rm mb}$, the local standard of rest velocity V$_{\rm LSR}$, and the integrated intensity I$({\rm C^{18}O}$). We computed the physical parameters of the clumps using the same methods and equations as detailed in \cite{Takekoshi2019}. In order to verify the actual existence of these clumps, we conducted a comparison with the FCRAO $^{13}$CO(J=2$\rightarrow$1) data reported by \cite{Schneider2006} and \cite{Schneider2011}. Furthermore, we use the definition provided by \cite{Yamagishi2018}, \cite{Saito2007}, \cite{Williams2000}, and references therein, which characterizes molecular condensations as having H$_2$ number densities on the order of $\sim 10^{4}-10^5$ cm$^{-3}$ and local thermodynamic equilibrium (LTE) masses ranging from 15 to 1500 M$_\odot$. This comparison serves to validate and confirm the nature of the identified clumps, ensuring that they meet the criteria of genuine molecular condensations in the observed region.

 %During the analysis, any clumps that contained one or more identified detections within the same spatial field were excluded from further consideration.

\section{Results and Discussion}
\label{sec:res_disc}

%In Fig.~\ref{fig:c18o_mom8_clumps} (bottom), we show the respective Schneider's maps (\cite{Schneider2006,Schneider2011}) in color, with our respective maps superimposed as contours.  We clearly identify 2 and 5 clumps in the DR-13S and DR-9 regions, in $^{13}$CO and C$^{18}$O observations respectively, but no C$^{18}$O clumps were detected in the DR-6 or DR-6W regions. Although this lack of detection can be explained by lack of sensitivity, we identified a set of 12 clumps in a molecular filament between DR-6 or DR-6W observed by Schneider. 

%Confirmamos que nuestros clumps tienen contraparte con Nikola.

%\begin{figure*}[!ht]
%\begin{center}
%\small
%\includegraphics[width=\columnwidth]{figs/cyg1_13co_spectra_all.pdf}
%\end{center}
%\caption{Spectra of $^{13}$CO for the 19 clumps reported in Tab.~\ref{tab:tab1}, fitted with a one-component 2-D Gaussian fit, shown as a red solid line.}
%\label{fig:spectra_13co}
%\end{figure*}

%\begin{figure*}[!ht]
%\begin{center}
%\small
%\includegraphics[width=\columnwidth]{figs/cyg1_c18o_spectra_all.pdf}
%\end{center}
%\caption{Spectra of C$^{18}$O for the 19 clumps reported in Tab.~\ref{tab:tab1}, fitted with a one-component 2-D Gaussian fit, shown as a red solid line.}
%\label{fig:spectra_c18o}
%\end{figure*}

\begin{figure*}[!ht]
\begin{center}
\includegraphics[width=\textwidth]{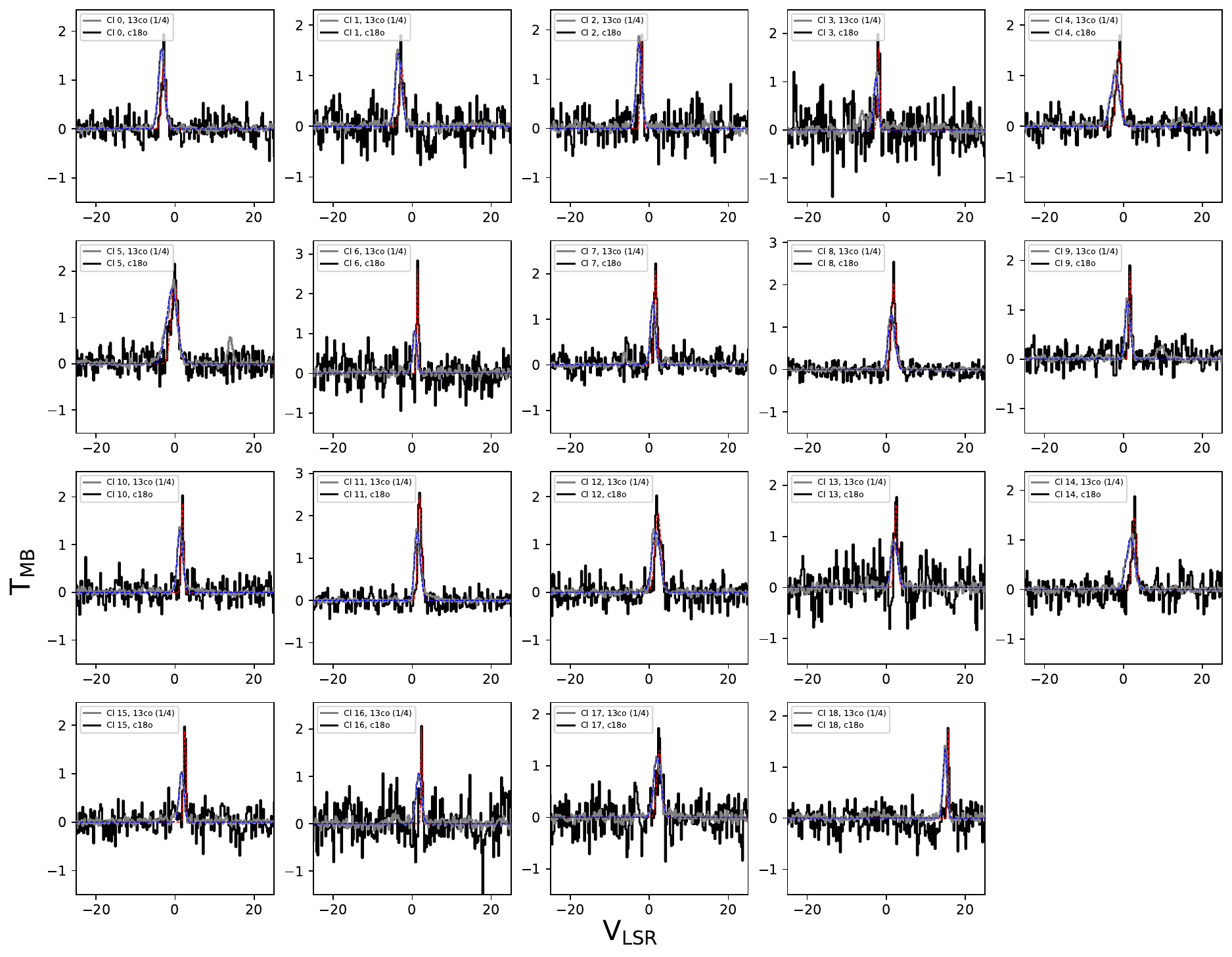}
\end{center}
\caption{Average spectra of $^{13}$CO and C$^{18}$O (J=1$\to$0) for the 19 clumps reported in Tab.~\ref{tab:tab1}. Gray and black solid curves correspond to the observed data of $^{13}$CO and C$^{18}$O, respectively. Blue and red dashed lines show the one-component Gaussian fit to the observed spectra of $^{13}$CO and C$^{18}$O, respectively.}
\label{fig:spectra_c18o}
\end{figure*}

 \begin{table}[!ht]
\centering
\tiny
\caption{Physical parameters of new C$^{18}$O clumps detected in DR-9, DR-13, and in a filament between DR-6 and DR-6W.}
\label{tab:tab1}
\begin{tabular}{cccccccccccc}

 \hline
Clump & l & b & V$_{\rm LSR}$ & $\Delta$V & T$_{\rm mb}^{\rm C^{18}O}$ &   T$_{\rm mb}^{\rm ^{12}CO}$ & I(${\rm C^{18}O}$) & R$_{\rm cl}$ & M$_{\rm LTE}$ & n$_{\rm H_2}$ & M$_{\rm vir}$ \\ 
{\tiny ID} & {\tiny ($\circ$)} & {\tiny ($\circ$)} & {\tiny (km s$^{-1}$)} & {\tiny (km s$^{-1}$)} & {\tiny (K)} & {\tiny (K)} & {\tiny (K km s$^{-1}$ pc$^{2}$)} & {\tiny (pc)} & {\tiny (10$^{2}$ M$_\odot$)} & {\tiny (10$^{4}$ cm$^{-3}$)} & {\tiny (10$^{2}$ M$_\odot$)} \\ \hline

0  &  77.9310 & -0.0402 &  -2.88$\pm$0.06 &  1.26$\pm$0.13 &   1.4$\pm$0.13 &  16.98$\pm$0.35 &  1.29$\pm$0.18 &  0.47 &  4.51$\pm$0.63 &  1.56$\pm$0.22 &  3.23$\pm$0.09 \\
1  &  77.9694 & -0.0459 &   -2.7$\pm$0.09 &   1.25$\pm$0.2 &  1.22$\pm$0.17 &  14.44$\pm$0.32 &   0.5$\pm$0.11 &  0.31 &   1.6$\pm$0.34 &   1.84$\pm$0.4 &   2.36$\pm$0.1 \\
2  &  77.9060 & -0.0025 &  -1.98$\pm$0.04 &  0.65$\pm$0.09 &  1.98$\pm$0.25 &  17.63$\pm$0.44 &  0.36$\pm$0.07 &  0.29 &  1.29$\pm$0.25 &  1.91$\pm$0.36 &  1.62$\pm$0.07 \\
3  &  78.0478 &  0.0213 &  -1.94$\pm$0.07 &  0.74$\pm$0.16 &  1.76$\pm$0.33 &   5.71$\pm$0.31 &   0.2$\pm$0.06 &  0.21 &  0.42$\pm$0.12 &  1.56$\pm$0.45 &  0.72$\pm$0.08 \\
4  &  78.0038 &  0.0339 &  -1.09$\pm$0.06 &  1.55$\pm$0.13 &  1.52$\pm$0.11 &   8.07$\pm$0.29 &  1.31$\pm$0.15 &  0.41 &  3.13$\pm$0.36 &  1.64$\pm$0.19 &  5.86$\pm$0.26 \\
5  &  78.1138 & -0.3035 &  -0.15$\pm$0.06 &  2.05$\pm$0.14 &   1.77$\pm$0.1 &    9.2$\pm$0.39 &  1.65$\pm$0.15 &  0.37 &  4.16$\pm$0.38 &  2.95$\pm$0.27 &   7.75$\pm$0.3 \\
6  &  77.8382 &  0.2567 &   1.38$\pm$0.03 &  0.54$\pm$0.08 &  2.64$\pm$0.32 &   8.91$\pm$0.28 &  0.25$\pm$0.05 &  0.23 &  0.63$\pm$0.12 &  1.84$\pm$0.34 &  0.77$\pm$0.07 \\
7  &  77.8078 &  0.2995 &   1.61$\pm$0.03 &  0.87$\pm$0.08 &  2.07$\pm$0.15 &   8.64$\pm$0.23 &  2.15$\pm$0.24 &  0.60 &   5.28$\pm$0.6 &   0.89$\pm$0.1 &  2.43$\pm$0.11 \\
8  &  77.7859 &  0.5438 &   1.82$\pm$0.03 &  1.26$\pm$0.07 &   2.06$\pm$0.1 &   9.14$\pm$0.17 &   1.7$\pm$0.12 &  0.44 &  4.29$\pm$0.31 &  1.75$\pm$0.13 &  3.59$\pm$0.11 \\
9  &  77.7618 &  0.3475 &   1.67$\pm$0.03 &  0.69$\pm$0.07 &   1.8$\pm$0.17 &   9.45$\pm$0.43 &   0.69$\pm$0.1 &  0.41 &  1.77$\pm$0.25 &  0.93$\pm$0.13 &  1.96$\pm$0.11 \\
10 &  77.7898 &  0.3982 &   1.83$\pm$0.03 &  0.72$\pm$0.08 &  1.89$\pm$0.17 &   9.54$\pm$0.21 &  0.48$\pm$0.07 &  0.33 &  1.24$\pm$0.18 &  1.28$\pm$0.18 &  1.81$\pm$0.08 \\
11 &  77.7850 &  0.4838 &    1.9$\pm$0.03 &  1.12$\pm$0.06 &  2.49$\pm$0.12 &   9.45$\pm$0.19 &   1.2$\pm$0.09 &  0.36 &  3.07$\pm$0.22 &  2.37$\pm$0.17 &   2.2$\pm$0.07 \\
12 &  77.7995 &  0.5995 &   2.22$\pm$0.06 &  1.78$\pm$0.15 &  1.66$\pm$0.12 &  10.22$\pm$0.19 &   0.97$\pm$0.1 &  0.31 &  2.56$\pm$0.28 &  2.96$\pm$0.32 &  3.81$\pm$0.15 \\
13 &  77.8288 &  0.6551 &   2.46$\pm$0.06 &  0.98$\pm$0.15 &  1.67$\pm$0.22 &   8.87$\pm$0.31 &  0.41$\pm$0.08 &  0.27 &  1.03$\pm$0.21 &  1.76$\pm$0.36 &  2.23$\pm$0.17 \\
14 &  77.8293 &  0.6125 &   2.74$\pm$0.06 &  1.27$\pm$0.13 &  1.49$\pm$0.13 &    8.65$\pm$0.2 &  0.67$\pm$0.09 &  0.33 &  1.65$\pm$0.23 &   1.7$\pm$0.24 &  4.22$\pm$0.19 \\
15 &  77.9109 &  0.6747 &   2.47$\pm$0.03 &  0.65$\pm$0.08 &    2.0$\pm$0.2 &   9.68$\pm$0.34 &   0.5$\pm$0.08 &  0.34 &   1.28$\pm$0.2 &  1.19$\pm$0.18 &   1.56$\pm$0.1 \\
16 &  77.8046 &  0.3415 &   2.41$\pm$0.05 &  0.39$\pm$0.09 &   2.1$\pm$0.44 &   8.57$\pm$0.42 &  0.14$\pm$0.04 &  0.23 &  0.35$\pm$0.11 &  1.03$\pm$0.32 &  1.29$\pm$0.13 \\
17 &  77.7820 &  0.6425 &    2.57$\pm$0.1 &  1.48$\pm$0.23 &  1.31$\pm$0.18 &   9.82$\pm$0.28 &  0.39$\pm$0.08 &  0.25 &  1.02$\pm$0.21 &  2.43$\pm$0.51 &  2.35$\pm$0.13 \\
18 &  78.0178 & -0.2475 &  15.68$\pm$0.03 &  0.43$\pm$0.07 &  1.82$\pm$0.27 &  14.13$\pm$0.44 &  0.22$\pm$0.05 &  0.29 &  0.68$\pm$0.16 &  1.02$\pm$0.23 &  0.82$\pm$0.05 \\
\hline

\end{tabular}

%{\tiny \raggedright $^{a}$ For the calculations of the physical parameters a distance of 1.6 $\pm$ 0.1 kpc was assumed. The angular size of the [FTK--MC] molecular cloud is considered to be 0.55 $\pm$ 0.02 degrees, which is the sum of the sizes of the two individual components [FTK--MC]HS and [FTK--MC]J2108.\\
%        $^{b}$For the virial mass calculation an average of the line-widths of C$_{1}$ of the $^{13}$CO($J$=1$\rightarrow$0) emission from [FTK--MC]HS and [FTK--MC]J2108 was used.  
%        \par}
       
\end{table}

In Fig.~\ref{fig:c18o_mom8_clumps} (top), we present our maps of $^{13}$CO(J=1$\rightarrow$0) and C$^{18}$O(J=1$\rightarrow$0). The identified gas condensations, which have been studied and recognized, are denoted by contours and labeled with corresponding numbers. In the bottom panels of Fig.~\ref{fig:c18o_mom8_clumps}, we present a comparison between a combination of NRO maps (shown as contours) and FCRAO $^{13}$CO(J=1$\to$0) observations, as reported by \cite{Schneider2011} (shown in colors). Every identified C$^{18}$O clump has a corresponding $^{13}$CO(J=1$\to$0) emission, which we can also match with the Schneider FCRAO data. Specifically, we have identified one clump associated with the DR-13S region, five clumps for DR-9, and 12 clumps in a filament located between the DR-6 and DR-6W regions, as previously observed by \cite{Schneider2006}. Interestingly, while the NRO-45m $^{13}$CO(J=1$\to$0) emission is detected in the DR-6 and DR-6W regions, no corresponding C$^{18}$O clumps were identified for both sources. In Fig.~\ref{fig:mosaic_12co_13co_c18o}, we present the complete Nobeyama Cygnus-X survey maps of $^{12,13}$CO (top) and C$^{18}$O (bottom left), complementing the studies by \cite{Takekoshi2019} and \cite{Yamagishi2018}. Furthermore, we include the FCRAO $^{13}$CO map of \cite{Schneider2006} and \cite{Schneider2011} (shown in colors), overlaid with the NRO 45-m $^{13}$CO emission as contours. The excellent morphology match between the NRO-45 m $^{13}$CO(J=1$\to$0) and FCRAO $^{13}$CO(J=1$\to$0) observations reported by \cite{Schneider2011} confirms the filamentary structure observed in this segment of the Cygnus-X region. In Fig.~\ref{fig:spectra_c18o} we present the spectra of $^{13}$CO and C$^{18}$O for the 19 identified C$^{18}$O clumps. The physical parameters for these clumps are provided in Tab.\ref{tab:tab1}. 

The identified C$^{18}$O clumps within the region covered by the observations exhibit an average radius of approximately $\sim$0.3 pc, H$_2$ densities of around $\sim$1.7$\times 10^{4}$ cm$^{-3}$, and LTE masses of approximately $\sim$10$^2$ M$_\odot$, which meet the definition of a molecular clump mentioned in Section \ref{sec:observations}. Additionally, the estimated physical parameters agree with other massive star formation regions examined by \cite{Saito2007}. Although the average radius and LTE masses align with the findings of \cite{Takekoshi2019}, the H$_2$ density is slightly higher, differing by a factor of 2. Nonetheless, it is essential to note that the calculated LTE masses do not reach values greater than 10$^3$ M$_\odot$, indicating that these clumps are unlikely to evolve into open clusters containing high-mass stars (>10$^8 \rm M_\odot$)\cite{Takekoshi2019}. However, to draw definitive conclusions regarding the evolution into open clusters, deeper observations are necessary.

\section{Conclusions}
\label{sec:conclusions}

Using new high-resolution (16'') observations obtained with the 45 m radio telescope at the Nobeyama Radio Observatory, we have conducted a study to complement the Nobeyama Cygnus-X survey, in particular covering the missing one square degree region encompassing the star-forming regions DR-6, DR-6W, DR-9, and DR-13S. %This complements the work done previously by \cite{Takekoshi2019} and \cite{Yamagishi2018}. 
In this study, we focused on $^{12,13}$CO(J=1$\rightarrow$0) and C$^{18}$O(J=1$\rightarrow$0) observations, and we successfully identified and analyzed the physical parameters of the detected C$^{18}$O clumps within this new field. The main results of our investigation are summarized as follows:

\begin{itemize}

    \item The molecular emission in the field shows a filamentary morphology and the dense gas, as traced by C$^{18}$O, appears compact, consistent with the findings of \cite{Takekoshi2019}.
    
    \item We have identified nineteen C$^{18}$O clumps: five for DR-9, one isolated, one for DR-13S (excluded in \cite{Takekoshi2019}), and 12 clumps for a filament between DR-6 and DR-6W, despite the fact that no clumps were detected for DR-6 and DR-6W. This discrepancy may be attributed to the sensitivity limitations of our observations.
    
    \item We have determined the physical parameters of these C$^{18}$O clumps, which exhibit molecular condensations with a radius of approximately 0.3 pc, H$_2$ densities of approximately 1.7$\times 10^{4}$ cm$^{-3}$, and LTE masses of about 10$^2$ M$\odot$. These parameters align with the operational definition of C$^{18}$O clumps observed in other regions of the southern part of the Nobeyama Cygnus-X Survey and other massive star-forming regions.

    \item The masses of the C$^{18}$O clumps are less than 5$\times$10$^2$ M${\odot}$, indicating that they lack the necessary mass ($>$ 10$^3$ M$_{\odot}$) to evolve into clusters hosting several OB stars, as discussed in \cite{Takekoshi2019}.
    
\end{itemize}

\noindent \textbf{\textit{Acknowledgments}}

 \small We thank an anonymous referee for comments and suggestions that improved the manuscript. EdelaF thanks the Inter-University Research Programme of the Institute for Cosmic Ray Research (ICRR), University of Tokyo (UTokyo), grant 2023i--F--005. IT--J gratefully acknowledges support from the Consejo Nacional de Ciencias y Tecnolog\'ia, M\'exico grant 754851. The authors thank the computational resources and technical support of the Centro de An\'alisis de Datos y Superc\'omputo of the Universidad de Guadalajara through the Leo-Atrox supercomputer. %The authors thank an anonymous referee for his reviews and valuable comments to improve the manuscript. 
 This research used astrodendro, a Python package to compute dendrograms of astronomical data (\url{http://www.dendrograms.org/}).

%% Full authors list (ONLY FOR COLLABORATIONS)
%\clearpage
%\section*{Full Authors List: \Coll\ Collaboration}
%
%\noindent \textbf{Note comment afterwards:} Collaborations have the possibility to provide an authors list in xml format which will be used while generating the DOI entries making the full authors list searchable in databases like Inspire HEP. For instructions please go to icrc2021.desy.de/proceedings or contact us under icrc2021proc@desy.de.\\
%
%\scriptsize
%\noindent
%first.author$^1$, 
%second.author$^2$, 
%third.author$^3$ % .... more names
%and 
%last.author$^{n}$ \\
%
%\noindent
%$^1$first.affiliation.
%$^2$second.affiliation. % .... more affiliation
%$^{m}$last.affiliation.

\end{document}